\def\year{2020}\relax
\documentclass[letterpaper]{article} 
\usepackage{aaai20}  
\usepackage{times}  
\usepackage{amsfonts}
 \usepackage{amsmath}
\usepackage{mathrsfs}
\usepackage{subfigure}
\usepackage{helvet} 
\usepackage{courier}  
\usepackage[hyphens]{url}  
\usepackage{graphicx} 
\usepackage{graphicx}  
\usepackage{color}  
\title{Learned Video Compression via Joint Spatial-Temporal Correlation Exploration }
\author{\Large \textbf{Haojie Liu\textsuperscript{\rm 1}, Han Shen\textsuperscript{\rm 2}, Lichao Huang\textsuperscript{\rm 2}, Ming Lu\textsuperscript{\rm 1}, Tong Chen\textsuperscript{\rm 1},  Zhan Ma\textsuperscript{\rm 1}\thanks{Zhan Ma is the corresponding author.}}\\ 
\textsuperscript{\rm 1} Nanjing University,\textsuperscript{\rm 2} Horizon Robotics\\ 
\{haojie, luming, tong\}@smail.nju.edu.cn, \{han.shen, lichao.huang\}@horizon.ai, mazhan@nju.edu.cn 
}
 
\begin{document}

\maketitle

\begin{abstract}
  Traditional video compression technologies have been developed over   decades in pursuit of higher coding efficiency. Efficient temporal information representation plays a key role in video coding. Thus, in this paper, we propose to exploit the temporal correlation using both first-order optical flow and second-order flow prediction. We suggest an one-stage learning approach to encapsulate flow as quantized features from consecutive frames which is then entropy coded with adaptive contexts conditioned on joint spatial-temporal priors to exploit second-order correlations. Joint priors are embedded in autoregressive spatial neighbors, co-located hyper elements and temporal neighbors using ConvLSTM recurrently.
  We evaluate our approach for the low-delay scenario with High-Efficiency Video Coding (H.265/HEVC), H.264/AVC and another learned video compression method, following the common test settings. Our work offers the state-of-the-art performance, with consistent gains across all popular test sequences. 
\end{abstract}

\section{Introduction}
Video content occupied more than 70\% Internet traffic, and it became a big challenge for transmission and storage along its explosive growth. Researchers, engineers, etc, continuously pursue the (next-generation) high-efficiency video compression for wider application enabling and larger market adoption. Conventional video compression approaches usually follow the  hybrid coding framework over decades~\cite{VideoCompressionHistoryReview} with hand-crafted tools for individual components. It is not efficient to jointly optimize the system in an end-to-end manner, especially for the inter tool efficiency exploration despite its great success of H.264/AVC \cite{wiegand2003overview} and H.265/HEVC \cite{sullivan2012overview}.

Recently, image compression algorithms \cite{balle2016end,li2017learning,balle2018variational,mentzer2018conditional,liu2019non} based on machine learning have shown great superiority in coding efficiency for {\it spatial redundancy} removal, compared with conventional codecs. These methods benefit from  non-linear transforms, deep neural network (DNN) based conditional entropy model, and joint rate-distortion optimization (RDO), under an end-to-end learning strategy. Learned video compression can be extended from image compression by further exploiting the {\it temporal redundancy or correlation}.
\begin{figure}[t]
     \centering
     \includegraphics[scale=0.11]{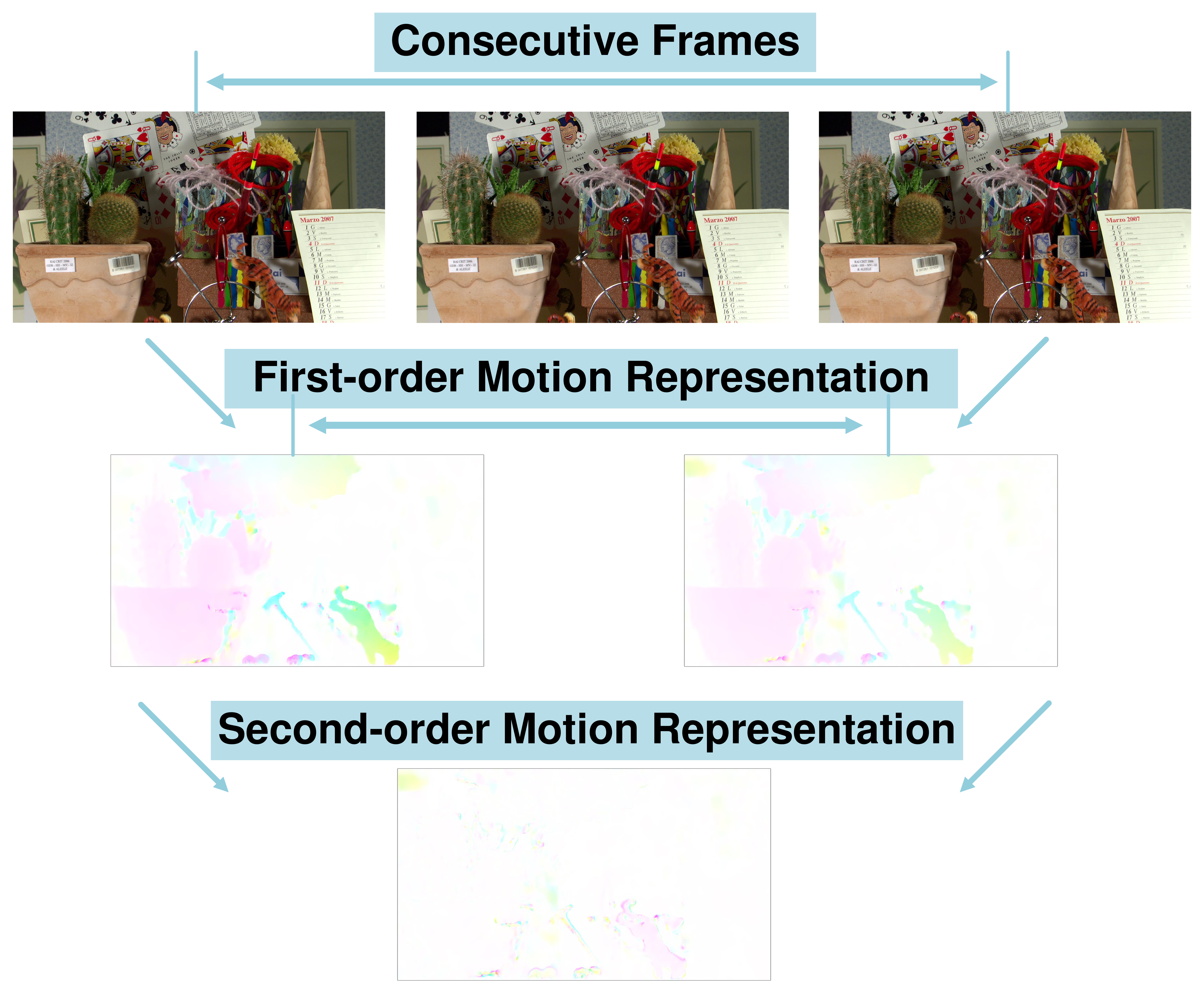}
     \caption{{\bf Temporal correlation exploration using both first-order and second-order motion representation.} Frame-to-frame redundancies can be easily removed by accurate flow estimation and compensation (e.g., first-order); Then flow-to-flow correlation can be further eliminated or reduced using predictions from joint spatial-temporal priors (e.g., second-order). }
     \label{fig:redundancies}
\end{figure}
We proposed an end-to-end video compression framework using joint spatial-temporal priors to generate  compact latent feature representations for intra texture, inter motion and sparse inter residual signals. Intra textures are well represented by spatial priors for both reconstruction and entropy context modeling using a variational autoencoder (VAE) structure \cite{minnen2018joint,liu2019non}.  We directly use NLAIC method proposed in~\cite{liu2019non} for our intra texture compression because of its state-of-the-art efficiency, and primarily investigate learned inter coding with the focus on efficient temporal motion representation in this paper.

We represent temporal information or correlation using its both first-order and second-order statistics.
The first-order temporal information is referred to as the motion fields (e.g., intensity, orientation) between consecutive frames. Motion fields can be described by either optical flow or block-based motion vectors. Here, we suggest an one-stage unsupervised flow learning approach, where first-order motions are quantized temporal features learned from the consecutive frames directly. Our unsupervised flow learning
does not rely on a well pre-trained optical flow estimation network, such as FlowNet2~\cite{ilg2017flownet,sun2018pwc}, and can derive the compressed optical flow from quantized features directly. 

The second-order temporal information is flow-to-flow correlations, describing the object acceleration.
Flow can be further predicted for energy reduction. Thus, we fuse priors from spatial, temporal and hyper
information to predict flow elements and compress the predictive difference. It turns out that this can be realized by entropy coding with adaptive contexts conditioned on fused priors.

Inter residual is derived between original frame and flow warped prediction. We reuse the intra texture coding network here directly for residual coding. It is worth to point out that VAE structures are applied for all components (e.g., intra, inter, residual) in this paper.


{\bf Contributions.}

1) We propose an end-to-end video compression method which offers the state-of-the-art performance against traditional video codecs and recent learned approaches with   consistent gains across a variety of common test sequences;

2) High-efficient inter coding is achieved by representing temporal correlation using both first-order optical flow and second-order flow predictive difference;

 3) First-order flow is offered using an one-stage unsupervised learning and represented by quantized features derived from consecutive frames;

 4) Second-order flow predictive difference compression is efficiently solved by entropy coding with adaptive contexts conditioned on fused priors (e.g., autoregressive spatial priors, hyperpriors, and temporal priors propagated using a ConvLSTM~\cite{xingjian2015convolutional}).



\section{Related Work}
\subsection{Learned Image Compression}
DNN based image compression approaches generally rely on autoencoders. \cite{TodericiVJHMSC16} first proposed to use recurrent autoencoders to progressively encode bits for image compression. Recent years, convolutional autoencoders are studied extensively, including non-linear transforms (e.g., generalized divisive normalization~\cite{balle2016end} and non-local attention transforms (NLAIC)~\cite{liu2019non}), differentiable quantization (e.g., soft-to-hard quantization~\cite{mentzer2018conditional} and uniform noise approximation~\cite{balle2016end}), and adaptive entropy model using the Bayesian generative rules (e.g., PixelCNNs~\cite{oord2016pixel} and variational autoencoders~\cite{balle2018variational}). RDO~\cite{Gary_RDO}
is applied by minimizing Lagrangian cost $J = R + \lambda D$,
in end-to-end training. Here, $R$ is referred to as entropy rate, and $D$ is the distortion measured
by either mean squared error (MSE) or multiscale structural similarity (MS-SSIM).

These approaches demonstrated better coding efficiency against traditional image coders, both objectively and subjectively. In addition, extreme compression are under exploration with adversarial training (e.g., conditional GANs \cite{agustsson2018generative}, multi-scale discriminators \cite{rippel2017real} for satisfied subjective quality at very low bit rates.
\subsection{Learned Video Compression}
Learned video compression~\cite{djelouah2019neural,habibian2019video} is a relatively new area. \cite{chen2017deepcoder} first proposed the DeepCoder where DNNs were used for intra texture and inter residual, and block motions were applied using traditional motion estimation for temporal information representation. Inspired by temporal interpolation and prediction \cite{niklaus2017video,niklaus2018context,NIPS2015_5778}, \cite{wu2018vcii} introduced a RNN based video compression framework through frame interpolation, offering comparable performance with H.264/AVC. However, interpolation typically brings structural delay.
Recently, unsupervised flow estimation methods~\cite{jason2016back,meister2018unflow} are introduced to utilize end-to-end learning for predicting optical flow between two frames without leveraging groundtruth flow. The proposed brightness constancy, motion smoothness and bidirectional census loss are proven to be efficient for flow generation.
Robust and reliable optical flow derivation methods are emerged, such as FlowNet2~\cite{ilg2017flownet} and PWC-Net~\cite{sun2018pwc}. FlowNet2 stacks multiple simple flownet and applies small displacements to correct flow. PWC-Net extends traditional pyramid flow reconstructed rules to learn and generate precise flow faster.

\cite{Lu_2019_CVPR} replaced the block based motion estimation with a pre-trained FlowNet2 followed by a cascaded autoencoder for flow compression. U-Net alike processing network was used to enhance the quality of predicted frame. In addition, they directly used models in \cite{balle2018variational} for their intra  and residual coding. The entire framework is so called DVC. DVC outperformed H.265/HEVC mainly at high bit rates but the coding efficiency dropped unexpectly at low bit rates as reported.

All these attempts in learned video compression were trying to represent the temporal information more efficiently.

\section{Proposed Method}
\subsection{Framework Overview}

\begin{figure}[t]
     \centering
     \includegraphics[scale=0.20]{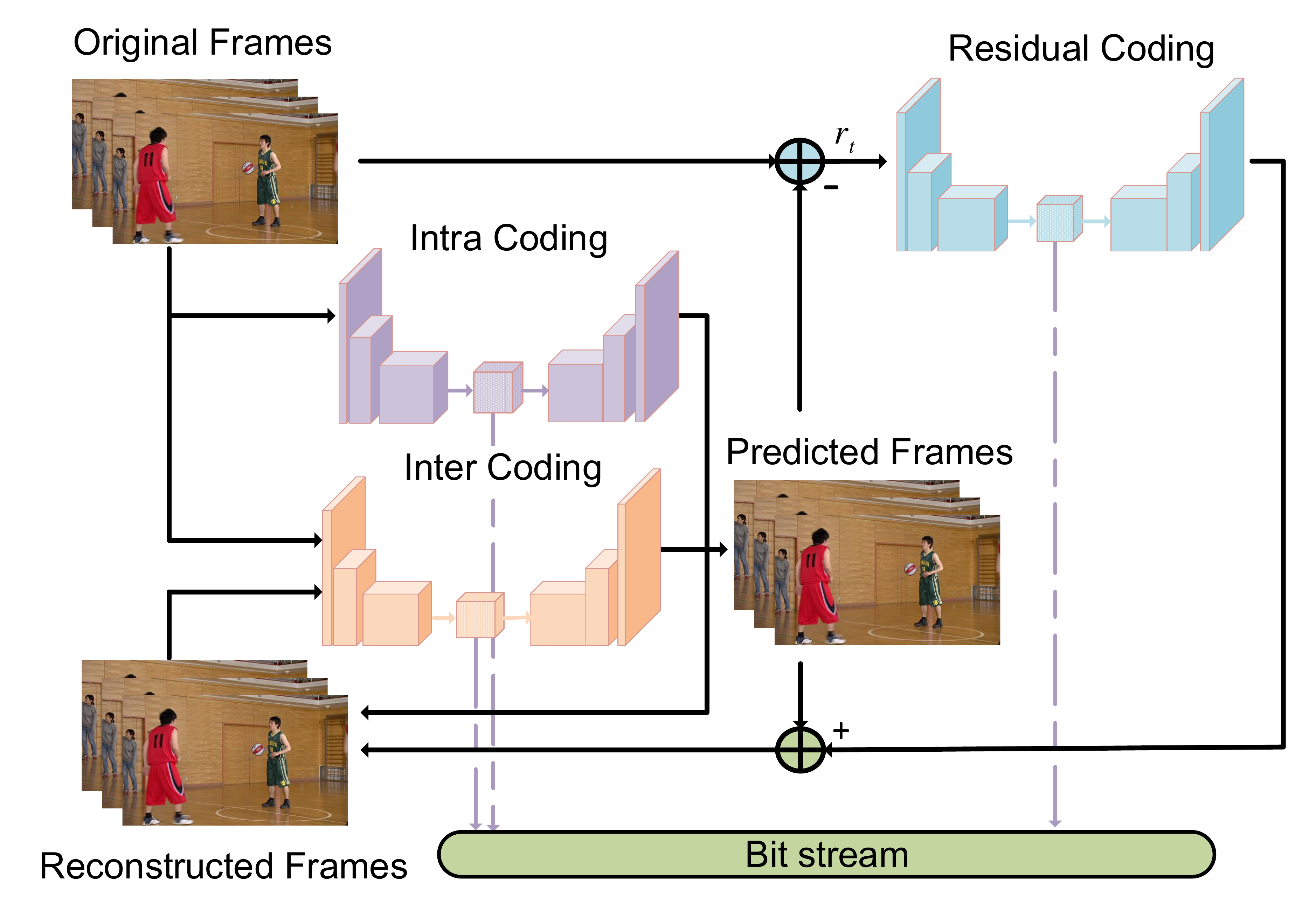}
     \caption{{\bf Learned Video Compression.} Intra \& residual coding using NLAIC in~\cite{liu2019non}, and inter coding by exploring one-stage unsupervised flow learning, and quantized flow feature encoding using context adaptive entropy models.}
     \label{fig:low_delay_video_compression}
\end{figure}

Fig.~\ref{fig:low_delay_video_compression} sketches our learned video compression. Given a group of pictures (GOP) $\mathbb{X}$ = \{${\bf X_1},{\bf X_2},...,{\bf X_t}$\}, we first encode ${\bf X_1}$ using NLAIC-based intra coding in~\cite{liu2019non}, having reconstructed frame as $\hat{\bf X}_1$.

For ${\bf X}_2$, we first learn the first-order flow for the predicted frame $\hat{\bf X}^p_2$ using $\hat{\bf X}_1$. Corresponding ${\bf r}_2 = {\bf X_2}-\hat{\bf X}^p_2$ is encoded using residual coding sharing the same architecture as NLAIC-based intra coding, and reconstructed as $\hat{\bf r}_2$. First-order flow is represented using quantized features that are entropy-coded with adaptive contexts conditioned on priors, by exploiting the second-order temporal correlation.
Final reconstruction $\hat{\bf X}_2$ is given by
${\hat{\bf X}_2} = {\hat{\bf X}^p_2} + {\hat{\bf r}_2}$. Subsequent frames follow the same process as of ${\bf X}_2$.

Adaptive entropy models for rate estimation and arithmetic encoding are embedded for intra, inter and residual components. Thanks to the VAE structures, intra and residual coding use the joint autoregressive spatial and hyper priors for the probability estimation; and inter coding applies priors from spatial, hyper and temporal information.

\subsection{Intra Coding} \label{sec:intra_coding}
\begin{figure}[t]
     \centering
     \includegraphics[scale=0.20]{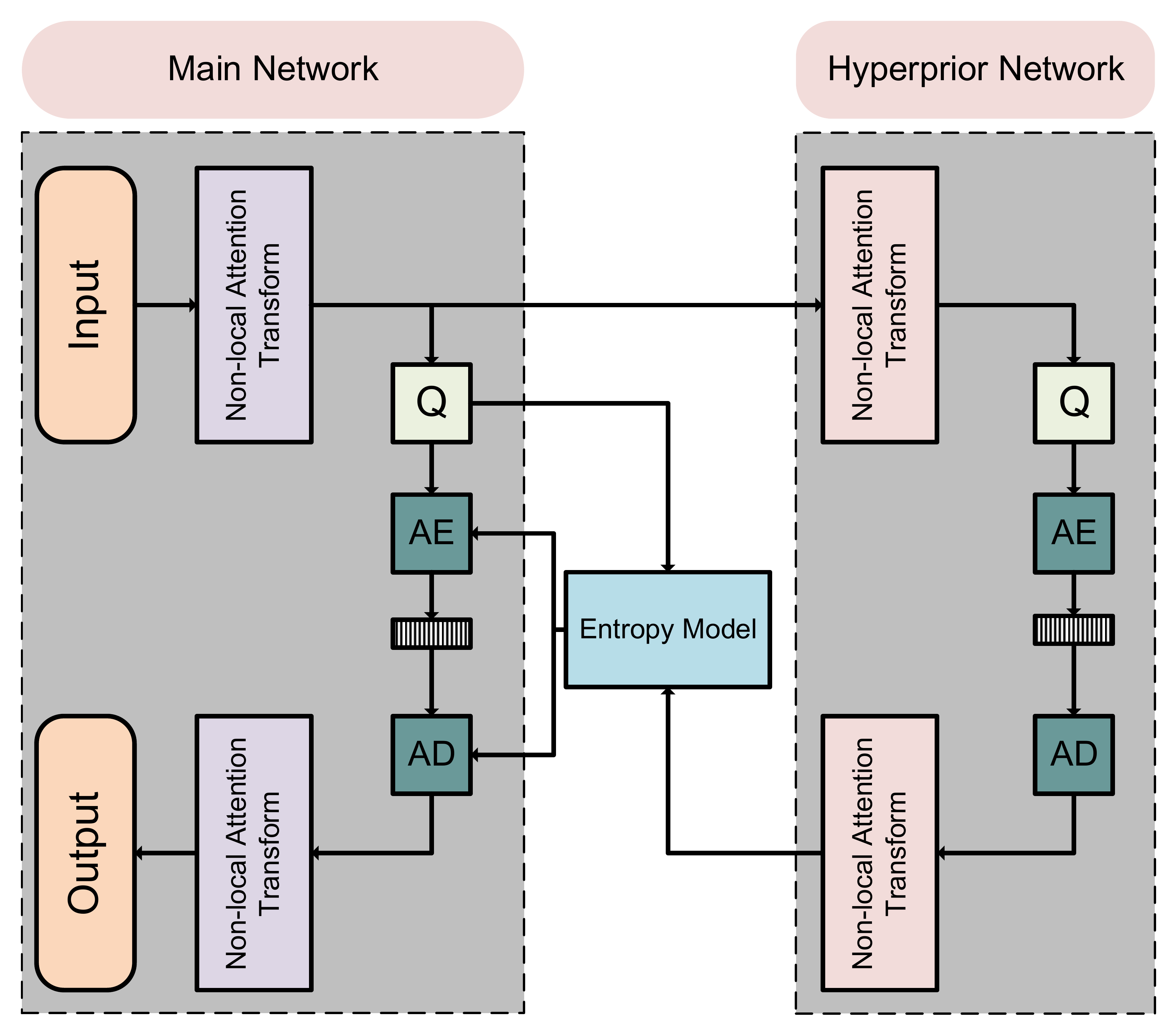}
     \caption{{\bf NLAIC-based Intra Coding.} A VAE architecture with non-local attention transforms embedded. ``AE'' represents the arithmetic encoding to encode the quantized features with  corresponding probability distribution, ``AD'' reverts binary strings to feature elements and ``Q'' is for quantization.}
     \label{fig:intra}
\end{figure}

We directly apply the NLAIC approach in~\cite{liu2019non} for our intra coding. Its state-of-the-art coding efficiency in image compression comes from the introduction of non-local attention transform that are embedded in both main and hyper encoder-decoder network in Fig.~\ref{fig:intra}. Note that the main network is used to obtain the reconstructed frame and the hyperprior network is used for context modeling of adaptive entropy coding.

In NLAIC method, non-local attention modules (NLAM) are embedded to capture joint local and global correlations for both reconstruction and context probability modeling, by inheriting the advantages from both nonlocal processing and attention mechanism. NLAM applies joint spatial-channel attention masks for more compact feature representation. And, masked 3D convolutions are used to fuse hyperpriors and autoregressive priors for accurate context estimation of adaptive entropy coding,
\begin{equation}
  ({\mu}_i,{\sigma}_i) = \mathbb F(\hat{x}_1,..., \hat{x}_{i-1},{\hat{\bf z}_t}).
  \label{eq:intra_feature}
\end{equation}
$\mathbb F$ represents cascaded 3D 1$\times$1$\times$1 convolutions to fuse the priors. $\hat{x}_1, {\hat {x}}_2,..., \hat{x}_{i-1}$ denote the causal (and possibly former reconstructed) pixels prior to current pixel $\hat{x}_i$ obtained by a 3D 5$\times$5$\times$5 masked convolution and $\hat{\bf z}_t$ are the hyperpriors. Probability of each pixel symbol in ${\bf \hat{x}_t}$ can be simply derived using
\begin{equation}
  p_{\hat{\bf x}_t|\hat{\bf z}_t}(\hat{\bf x}_t|\hat{\bf z}_t)  = {\prod_i} (\mathcal{N}{({\mu}_i,{\sigma}_i^2)} *\mathcal{U}(-\frac{1}{2},\frac{1}{2})) (\hat{x}_i),
  \label{eq:intra_texture_dist}
\end{equation} with a Gaussian distribution assumption with mean ($\mu$) and variance ($\sigma$).

\begin{figure*}[t]
\centering
\subfigure[]{\includegraphics[scale=0.20]{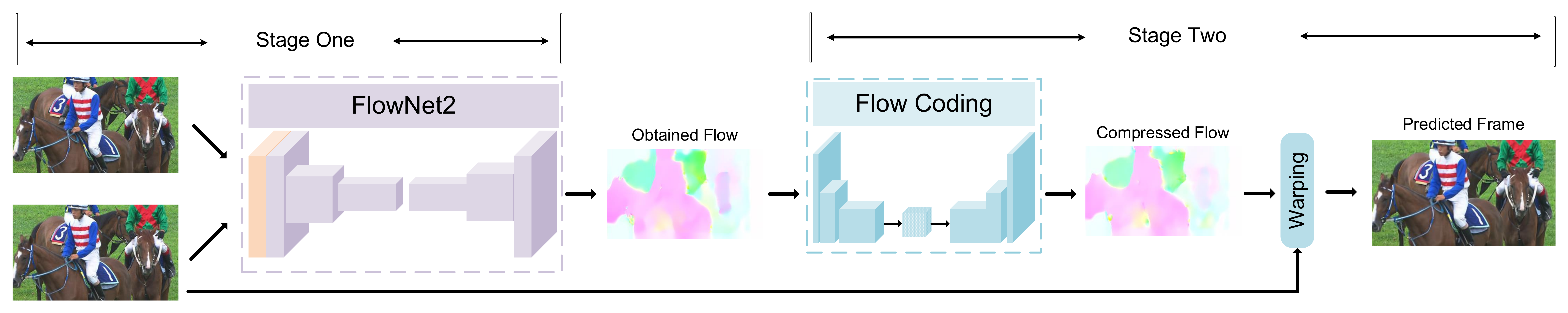} \label{sfig:two_stage}}
\subfigure[]{\includegraphics[scale=0.21]{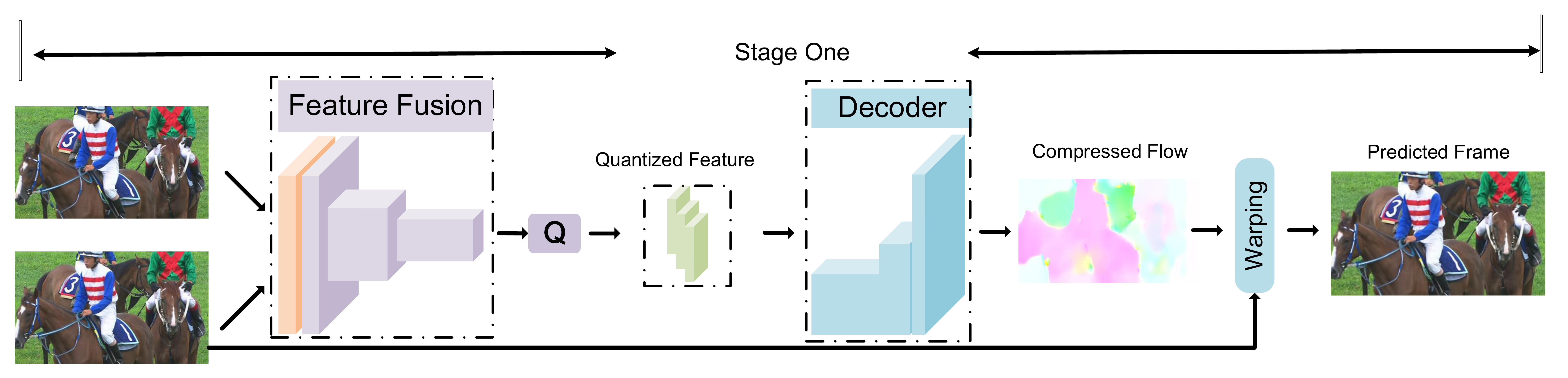} \label{sfig:one_stage}}
\caption{{\bf Flow Learning and Compensation.} (a) Two-stage supervised approach using a pre-trained flow net (with explicit raw flow) and  a cascaded flow compression autoencoder; (b) One-stage unsupervised approach with implicit flow represented by quantized features that will be directly decoded for compensation.}
\label{fig:flow_coding}
\end{figure*}

 \subsection{Inter Coding} \label{sec:inter_coding}

Video coding performance heavily relies on the efficient temporal information representation. It has two folds. One needs to have the most accurate first-order flow for compensation, and the other is to devise second-order statistics for flow prediction.

 \subsubsection{One-stage Unsupervised Flow Learning} \label{ssec:flow_learning}

Previous work in~\cite{Lu_2019_CVPR} obtained decoded optical flow using typical two-stage methods shown in Fig.~\ref{sfig:two_stage}. It relied on a well pre-trained flow network to  generate an uncompressed optical flow that was then compressed using a cascaded autoencoder. But, in our work, we leverage the quantized features between consecutive frames as compact motion representations and directly decode the compressed features for subsequent compensation. There is no need to explicitly derive raw and uncompressed flow in encoding process with supervised guidance as in FlowNet2 and PWC-Net. Thus, it is an one-stage unsupervised flow learning and compensation approach in Fig.~\ref{sfig:one_stage}.

We concatenated two consecutive frames as the input for feature fusion, and the network is consisted of stacked NLAM\footnote{Nonlocal attention is used to capture local and global correlations.} and downsampling (e.g., 5$\times$5 convolutions with stride 2), generating the fused features with $(H/16)\times(W/16)\times 64$ dimension. Quantization is then applied to obtain the quantized features $\mathscr{F}$ for entropy coding. The decoder mirrors the stacked NLAM with upsampling (e.g., 5$\times$5 deconvolutions with stride 2) in the feature fusion network, and derived the decoded flow $\hat{f}^d_t$ (at a size of $H \times W \times 2$ for separable horizontal and vertical orientations) for compensation. Here, $H, W$ denotes the height and width of the original frame, respectively.

To avoid quantization induced motion noise,we first pre-train the network with uncompressed consecutive frames ${\bf X}_{t-1}$ and ${\bf X}_t$. Then we replace ${\bf X}_{t-1}$  using its  decoded correspondence  ${\hat{\bf X}}_{t-1}$ as described in Eq.~(\ref{eq:decoded_flow}) and (\ref{eq:warped_frame}). Note that we only have ${{\hat{\bf X}}_{t-1}}$ not ${\bf X}_{t-1}$ for inter coding in practice. And we have directly utilized the decoded  flow $\hat{f}^d_t$ for end-to-end training and do not need a flow explicitly at encoding (i.e., implicitly represented by quantized features).

 A compressed flow representation of $\hat{f}^d_t$, i.e., quantized features $\mathscr{F}$, is encoded into the bitstream for delivery. $\hat{f}^d_t$ is then used for warping with reference frame to have $\hat{\bf X}^p_t$ for compensation, i.e.,
\begin{equation}
  \hat{f}^d_t = {\mathbb F}_d({\mathbb F}_e(\hat{\bf X}_{t-1},{\bf X}_t)), \label{eq:decoded_flow}
\end{equation}
\begin{equation}
  \hat{\bf X}^p_t = {\bf warping}(\hat{\bf X}_{t-1},\hat{f}^d_t ), \label{eq:warped_frame}
\end{equation}

Here ${\mathbb F}_e$ and ${\mathbb F}_d$ represent the feature fusion network with quantization and decoder network, respectively. Note that $\mathscr{F} = {\mathbb F}_e(\hat{\bf X}_{t-1},{\bf X}_t)$.

\subsubsection{Context Adaptive Flow Compression} \label{ssecion:flow_coding}
Previous section~\ref{ssec:flow_learning} introduced one-stage unsupervised flow learning targeting for the accurate first-order motion representation for compensation, where flow is represented implicitly using quantized features $\mathscr{F}$. Efficient representation of $\mathscr{F}$ is highly desirable

Generally, as shown in Fig.~\ref{fig:redundancies}, there is not only the first-order correlation (frame-to-frame) that can be exploited by the flow, but also the second-order correlation both spatially and temporally. A way to predict flow efficiently could lead to much better compression performance. Ideally, flow element can be estimated by its spatial neighbors, temporal co-located element, and hyper priors for energy compaction. A duality problem for such flow prediction using neighbors, is flow element entropy coding using adaptive contexts.

\begin{figure}[t]
     \centering
     \includegraphics[scale=0.26]{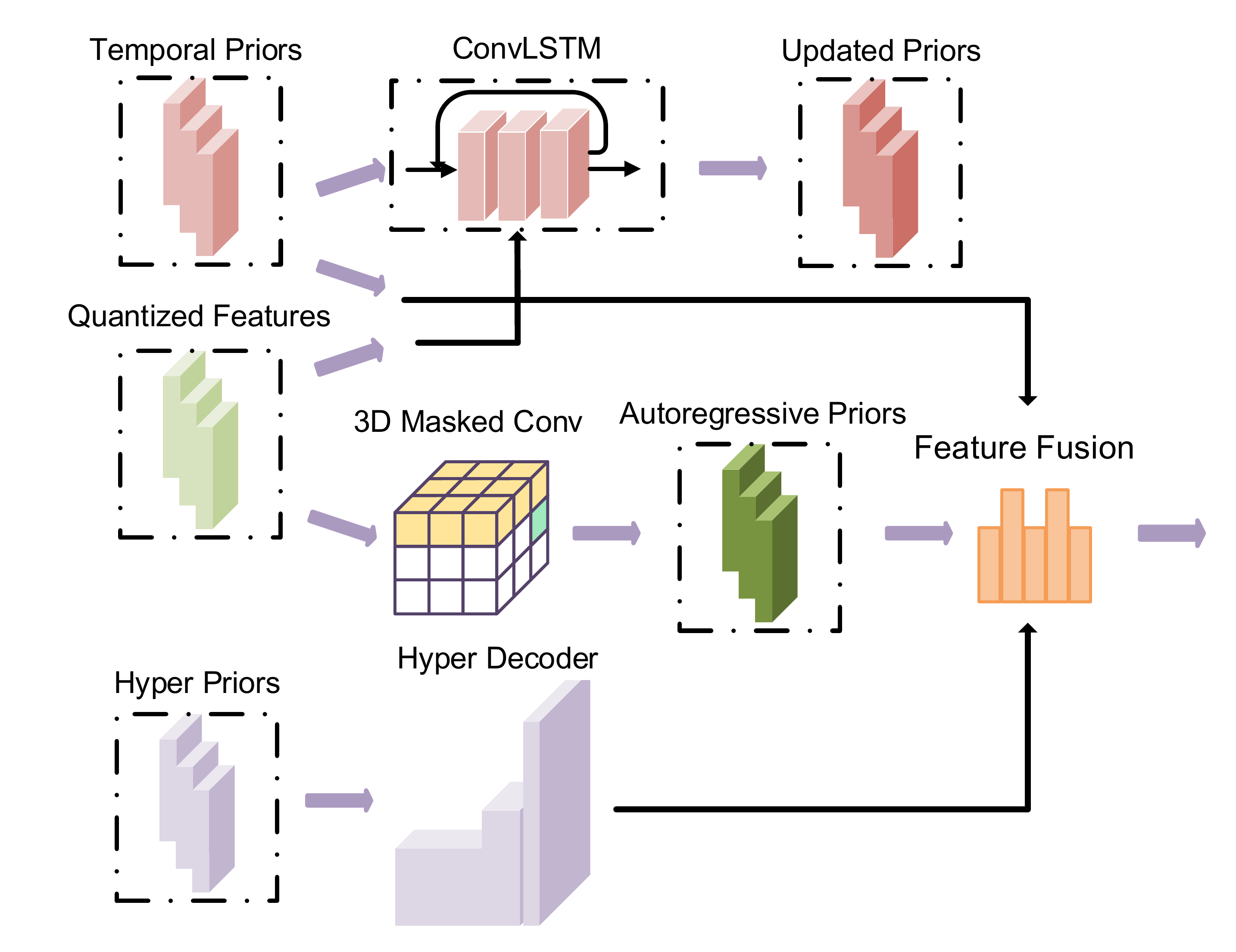}
     \caption{{\bf Adaptive Contexts Using Fused Priors.} Spatial autoregressive priors are applied using 3D masked convolutions, temporal priors are propagated using convLSTM, and hyperpriors are hyper decoder in a VAE setup. All priors are then fused together for probability modeling. Note that contexts update at a frame recurrent fashion.}
     \label{fig:entropy_model}
\end{figure}

Adaptive context modeling can leverage priors from spatial autoregressive neighbors, temporal and hyper information, shown in Fig.~\ref{fig:entropy_model}.  For spatial autoregressive prior, we propose to apply the 3D masked convolutions on quantized features $\mathscr{F}$; while for hyper priors, hyper decoder is used to decode corresponding information. Hyperpriors are widely used in VAE structured compression approaches. Note that, temporal correlations are exhibited in video sequence. Instead of applying the only pixel domain frame buffer in traditional video codecs, we propose to embed and propagate flow representation at a frame recurrent way using the ConvLSTM, which is also referred to as the temporal prior buffer. Temporal priors have the same dimension as the current quantized features $\mathscr{F}$.



These priors are fused together for context adaptive flow coding, i.e.,
\begin{equation}
  (\mu_{\mathscr{F}},\sigma_{\mathscr{F}}) = \mathbb F( {\mathscr{F}}_1,..., {\mathscr{F}}_{i-1}, \hat{\bf z}_t, {\bf h}_{t-1}), \label{eq:flow_probability}
\end{equation}
\begin{align}
  &p_{{\mathscr{F}}|({\mathscr{F}}_1,..., {\mathscr{F}}{i-1},\hat{\bf z}_t, {\bf h}_{t-1})}({\mathscr{F}}_i|{\mathscr{F}}_1,..., {\mathscr{F}}_{i-1},\hat{\bf z}_t, {\bf h}_{t-1}) \nonumber\\
  & \mbox{~~~~~~~~~~~~~~~~}=  {\prod_i} (\mathcal{N}{(\mu_{\mathscr{F}},\sigma_{\mathscr{F}}^2)} *\mathcal{U}(-\frac{1}{2},\frac{1}{2})) ({\mathscr{F}}_i).
  \label{eq:residual_dist}
\end{align}  $\mathscr{F}_i, i = 0, 1, 2, ...$ are elements of quantized features for implicit flow $\hat{f}^d_t$ representation, ${\bf h}_{t-1}$ is aggregated temporal priors from previous flow representations, which is updated using standard ConvLSTM:
\begin{equation}
  ({\bf h}_t, {\bf c}_t) = {\rm ConvLSTM}({\mathscr{F}_t, {\bf h}_{t-1}, {\bf c}_{t-1}}),
  \label{eq:trn_priors}
\end{equation}
where ${\bf h}_t, {\bf c}_t$ are updated state at $t$ and prepared for the next time step slot with ${\bf c}_{t-1}$  as a memory gate.

\subsection{Residual Coding} \label{ssec:residual_coding}
For the sake of simplicity, we encode the residual signals ${\bf r}_t$ using the identical networks as the NLAIC-based intra coding.  ${\bf r}_t$ is obtained by ${\bf r}_t = {\bf X}_t - {\bf X}^p_t$. Here, we do not calculate the loss between ${\bf r}_t$ and $\hat{\bf r}_t$ but directly target for overall reconstruction loss $\mathbb{D}({\bf X}_t,{\bf X}^p_t+{\hat{\bf r}_t})$ for optimizing the residual coding.

\section{Experimental Studies} \label{sec:exp}
We proceed to the details about training strategy and evaluation in this section. More ablation studies are given to verify the effectiveness of our work.
\begin{figure}[t]
\centering
\subfigure{\includegraphics[scale=0.26]{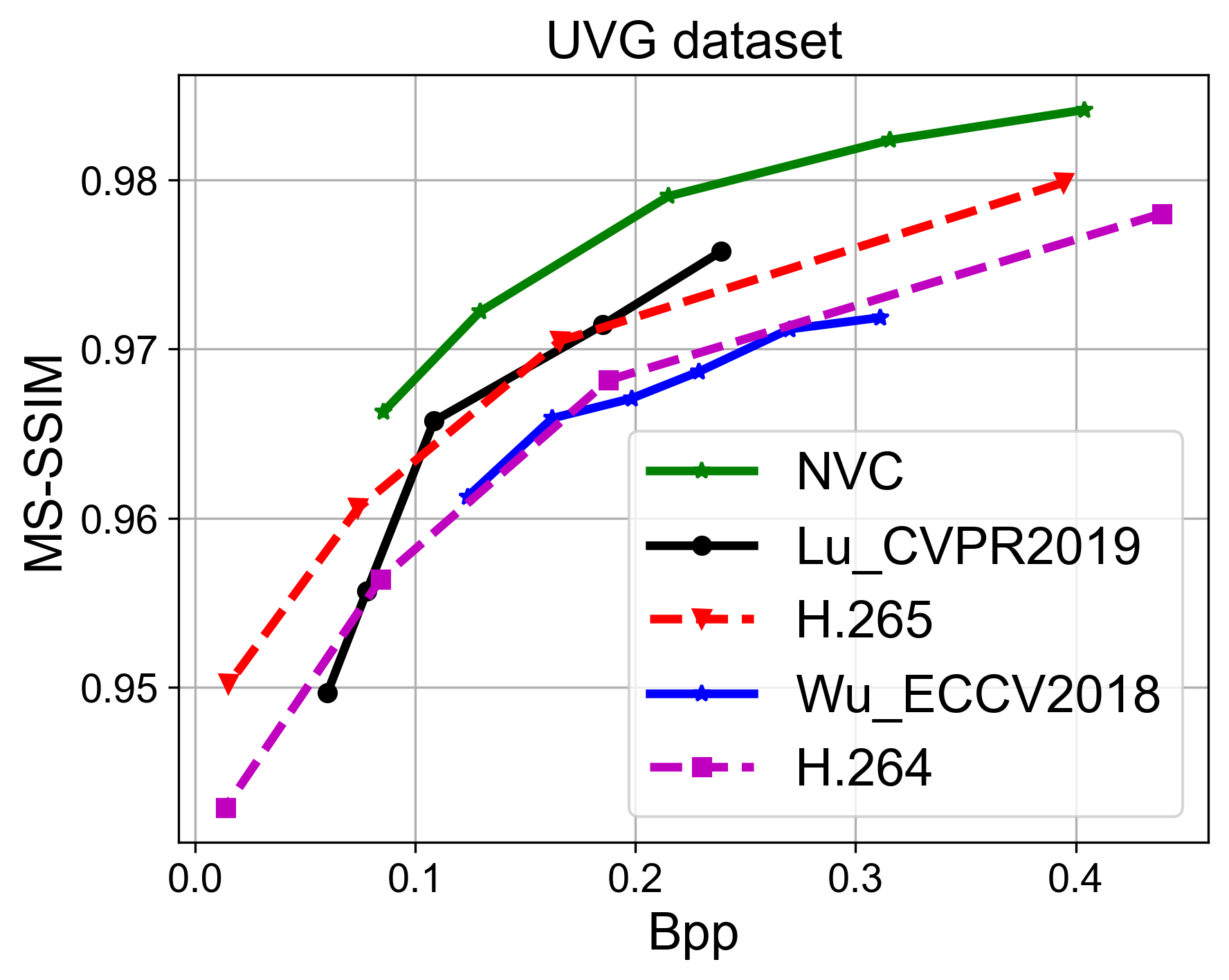}}
\subfigure{\includegraphics[scale=0.26]{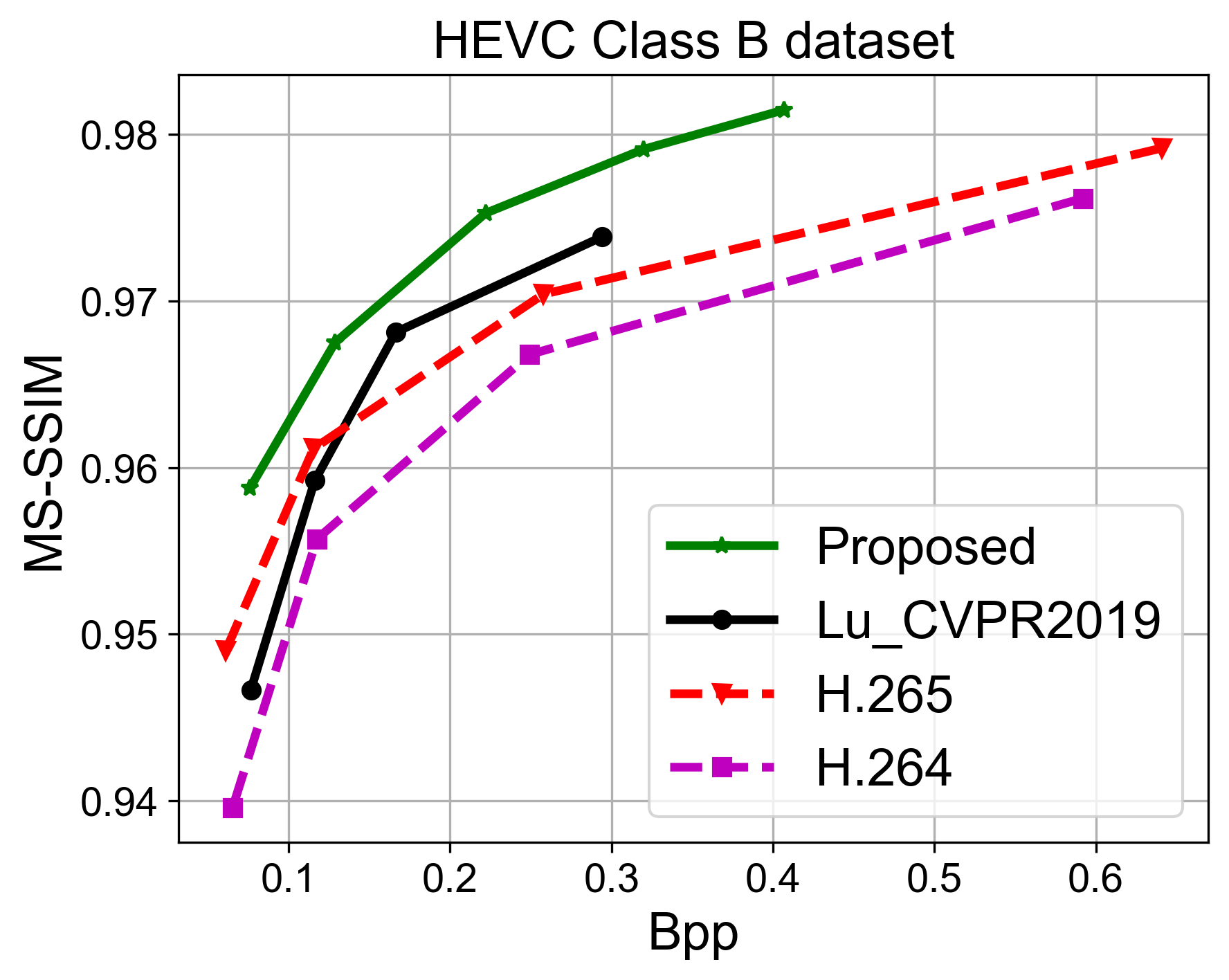}}\\
\subfigure{\includegraphics[scale=0.26]{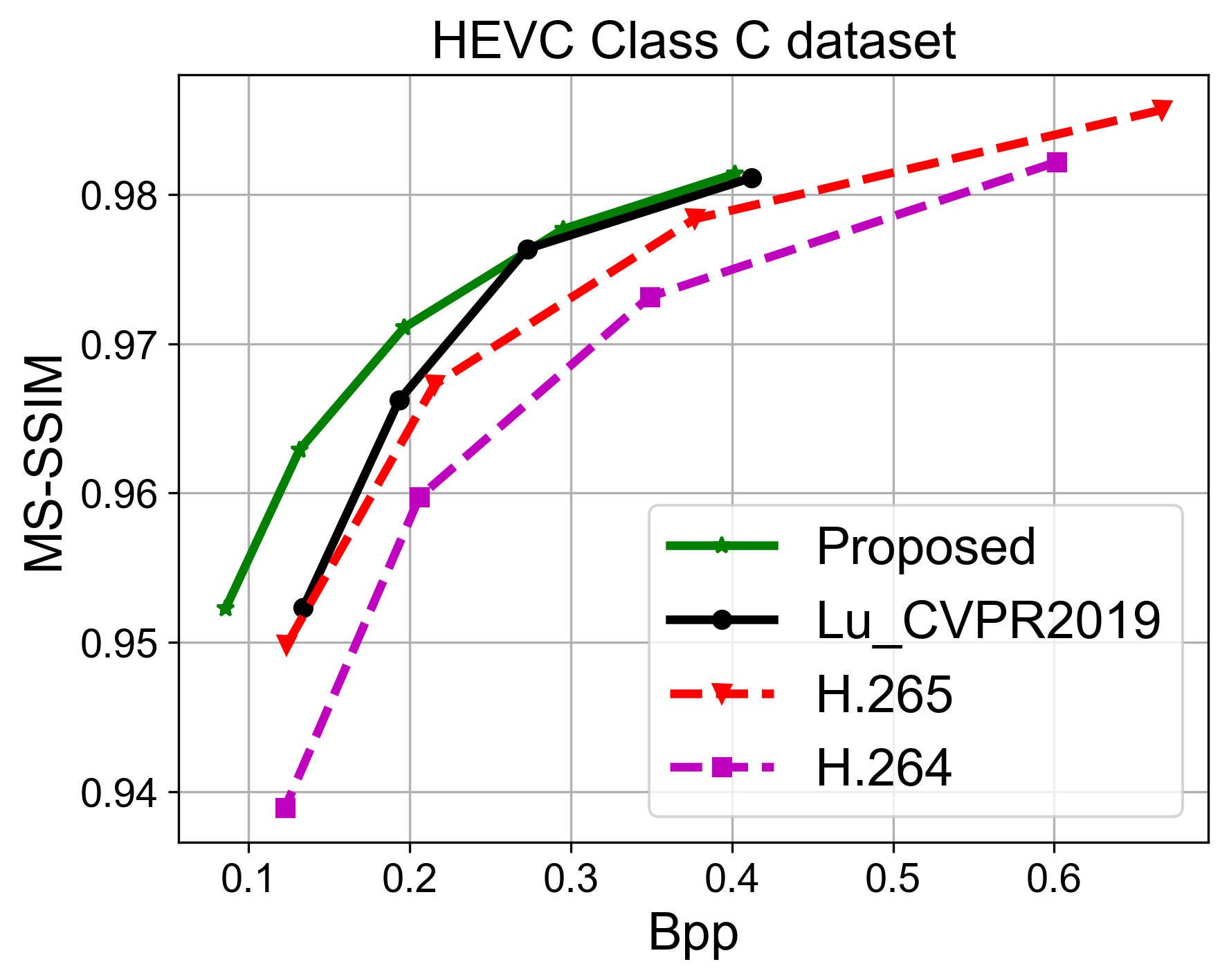}}
\subfigure{\includegraphics[scale=0.26]{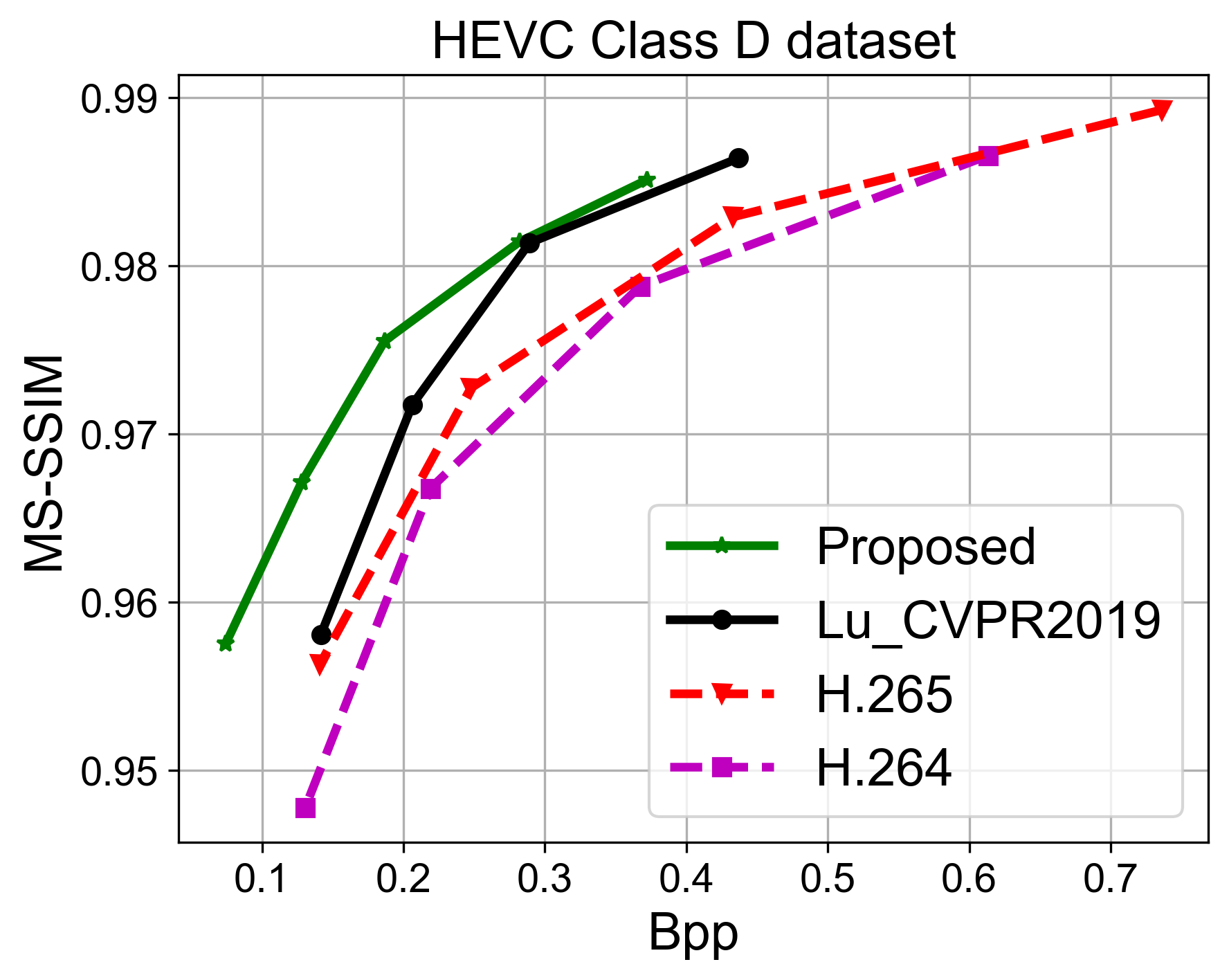}}\\
\subfigure{\includegraphics[scale=0.26]{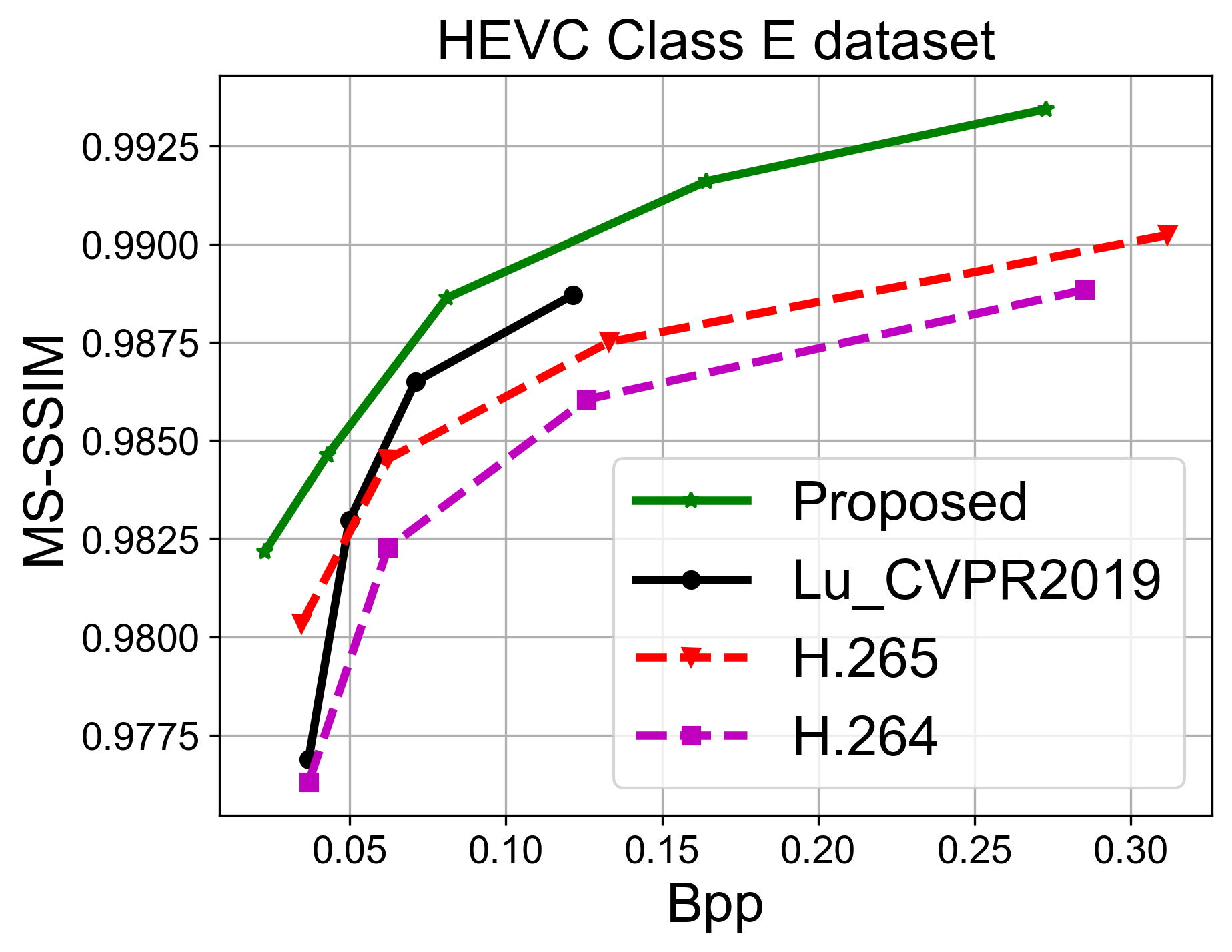}}
\subfigure{\includegraphics[scale=0.26]{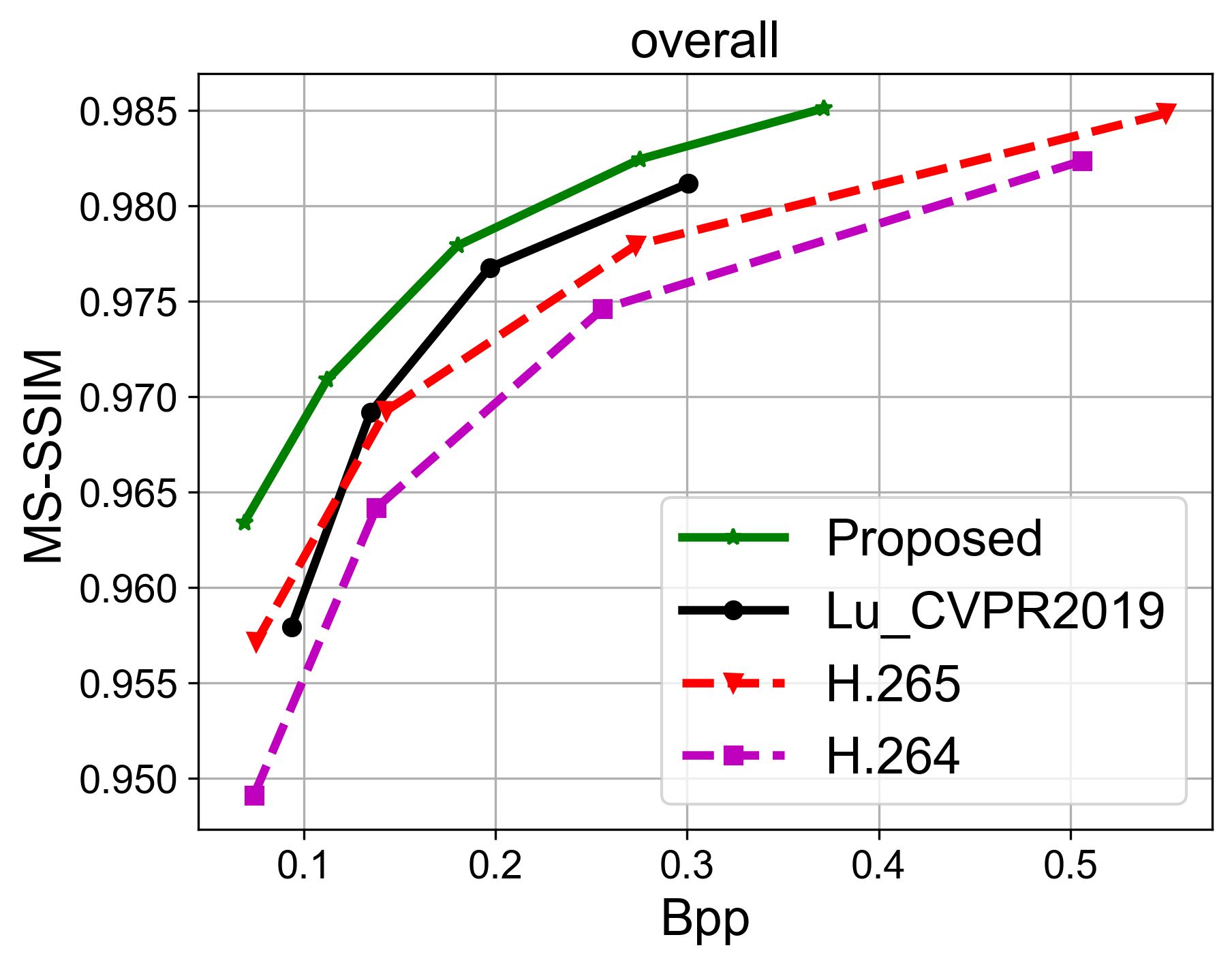}}
\caption{{\bf Rate-Distortion Performance Comparison.} Experiments are performance for HEVC comment test classes and UVG videos for a variety of content distributions. Our method gains consistently across all test videos.
}
\label{rd_curve}
\end{figure}

\subsection{Implementation Details}

{\bf Training \& Testing Datasets.} We choose COCO \cite{lin2014microsoft} dataset to pre-train NLAIC-based intra coding network. And then we joint train video compression framework on Vimeo 90k \cite{xue2019video} which is a widely used dataset for low-level video processing tasks. All images from these the datasets are randomly cropped into 192{$\times$}192 patches for training.

We gave evaluations on NLAIC using Kodak dataset in ablation studies to understand the efficiency of nonlocal attention transforms. We then evaluated our video compression approaches on standard HEVC dataset and ultra video group (UVG) dataset with different classes, resolution and frame rate.

{\bf Loss Function \& Training Strategy.} It is difficult to train multiple networks on-the-fly at one shot. Thus, we pre-train the intra coding and flow learning and coding networks first, followed by the jointly training with pre-trained network models for an overall optimization, i.e.,

\begin{align}
  L =  {\frac{\lambda_1}{n}{\sum_{t=0}^{n}}{\mathbb D}_1(\hat{{\bf X}}_t,{{\bf X}_t})}
     &+ {\frac{\lambda_2}{n}{\sum_{t=0}^{n}}{\mathbb D}_2(\hat{\bf X}_t^{p},{{\bf X}_t})} \nonumber\\
     &+  R_s+{\frac{1}{n-1}{\sum_{t=1}^{n}}R_t}, \label{eq:rdo}
\end{align} where ${\mathbb D}_1$ is measured using MS-SSIM, and ${\mathbb D}_2$ is the warping loss evaluated using $L$1 norm and total variation loss. $R_s$ represents the bit rate of intra frame and $R_t$ is the bit rate of inter frames including bits for residual and flow respectively. Currently, $\lambda_1$ and $\lambda_2$ will be adapted according to the specified overall bit rate and bit consumption percentage of flow information in inter coding.

Besides, entropy rate loss $R$ is approximated by conditional probability $p$ using Eq.~(\ref{rate_loss}), with main payload for context adaptive feature elements, and payload overhead for image height ($H$), width ($W$), number of frames ($N$) and GOP length ($n$), e.g.,
\begin{equation}
  R = - {\sum\nolimits_i} {\log_2}(p)+overhead.
   \label{rate_loss}
\end{equation} Note that bit consumption for ``overhead" is less than 0.1\% of the entire compressed bitstream, according to our extensive simulations.

To well balance the efficiency of temporal information learning and training memory consumption, we have enrolled 5 frames to train the video compression framework and shared the weights for the subsequent frames. The initial learning rate (LR) is set to 10e-4 and is clipped by half for every 10 epochs. The final models are obtained using a LR of 10e-5. We apply the distributed training on 4 GPUs (Titan Xp) for 5 days.

{\bf Evaluation Criteria.}
For fair comparison, we have applied the same setting as DVC in~\cite{Lu_2019_CVPR} for our method and traditional H.264/AVC, and HEVC codecs. We use GOP of 10 and encode 100 frames on HEVC test sequences and use GOP of 12 with 600 frames on UVG dataset. The reconstructed quality are measured in RGB domain using MS-SSIM. Bits per pixel (Bpp) is used for bit rate measure which can be easily translated to the {\tt kbps} by scaling it with $H\times W\times n/\tau$. $\tau$ is the video duration in seconds.

\subsection{Performance Comparison}

{\bf Rate distortion Performance.} Our approach outperforms all the existing methods as shown in Fig.\ref{rd_curve}.  Here, the distortion is measured by MS-SSIM which is proven to be a more relevant to human visual system, and used widely in learned compression methodologies~\cite{minnen2018joint}.

To the best of our knowledge, our work is the first end-to-end method that outperforms H.265/HEVC consistently across a variety of bit rates for all test sequences. In contrast, algorithm in
\cite{wu2018vcii}  only presents a similar performance as H.264/AVC. DVC~\cite{Lu_2019_CVPR} improves \cite{wu2018vcii} with better coding efficiency against HEVC at high bit rates. However, a cliff fall of performance is revealed for DVC at low bit rate (e.g., some rates having performance even worse than H.264/AVC). We have also observed that DVC's performance varies for different test sequences. But, our approach shows consistent gains, across contents and bit rates, leading to the conclusion that our model presents better generalization for practical applications.

We use H.264/AVC as the anchor for BD-Rate calculation as shown in Table~\ref{tab:BDrate}. Our approach reports 50.36\% and 51.67\% BD-rate reduction on HEVC test sequences and UVG dataset compared with H.264/AVC, respectively, offering a significant performance improvement margin in contrast to the HEVC or DVC over the H.264/AVC.

%

 \begin{figure}[h]
      \centering
      \includegraphics[scale=0.33]{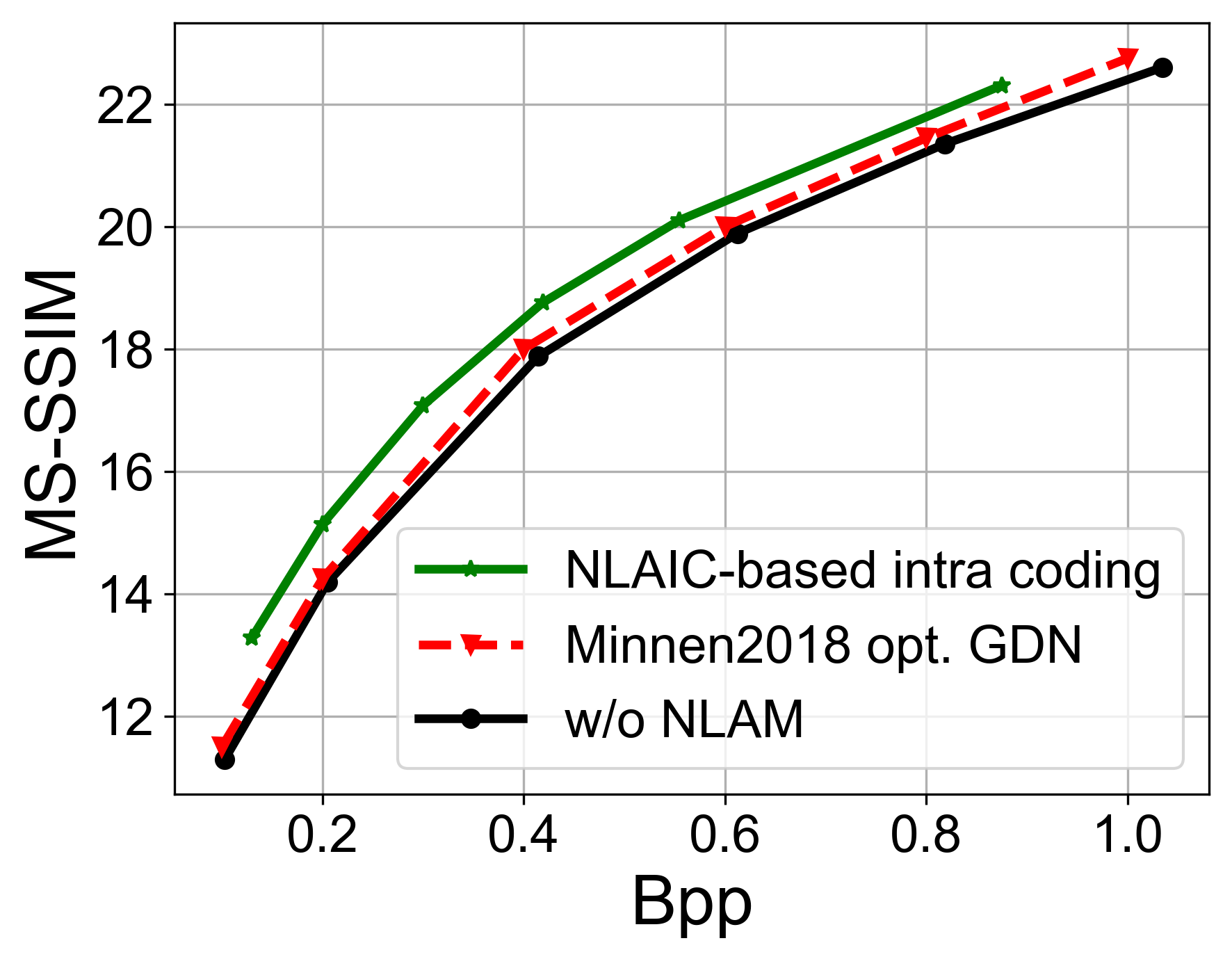}
      \caption{{\bf Efficiency of NLAM.} NLAM is used in non-local attention transforms for intra \& residual coding. Performance is reduced by removing NLAM, but still close to the work in~\cite{minnen2018joint}.}
      \label{fig:nlam}
 \end{figure}

 \begin{table}[t]
  \centering
  \caption{BD-Rate Gain of Our Method, HEVC and DVC against the H.264/AVC}

  \begin{tabular}{|c|c|c|c|}
    \hline
       Sequences & H.265/HEVC & DVC &  Ours \\
    \hline
     ClassB & -28.31\% &  -29.09\% & -54.17\%\\
     ClassC & -20.50\% & -28.11\% & -39.17\%\\
     ClassD & -8.89\% & -27.35\% & -44.84\%  \\
     ClassE & -29.52\% & -33.91\% & -63.28\% \\
    \hline
     Average & -21.73\% & -29.73\%&  {\bf -50.36\%}\\
     \hline
     UVG dataset & -37.25\% & -28.28\%&  {\bf -51.67\%}\\
    \hline
  \end{tabular}
  \label{tab:BDrate}
\end{table}
\begin{figure*}[h]
     \centering
     \includegraphics[scale=0.073]{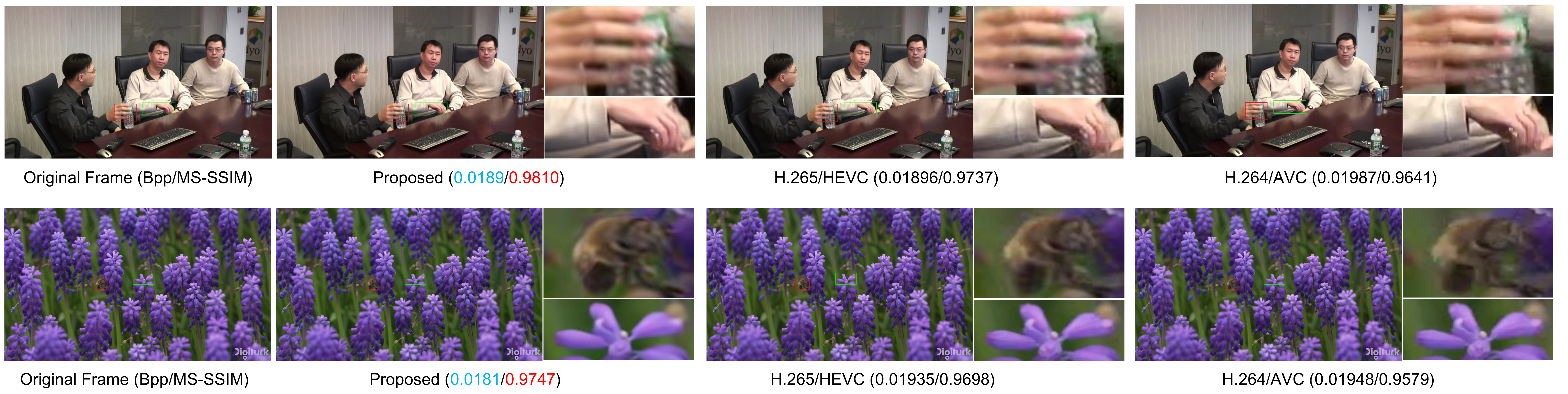}
     \caption{{\bf Visual Comparison.} Reconstructed frames of our method, H.265/HEVC and H.264/AVC. We avoid blocky artifacts and provide better quality of reconstructed frame at low bit rate.}
     \label{fig:visual_comparison}
\end{figure*}
{\bf Visual Comparison}
We provide the visual quality comparison with H.264/AVC and H.265/HEVC as shown in Fig.~\ref{fig:visual_comparison}.

 Traditional codecs usually suffer from blocky artifacts, especially at low bit rate, because of its block based coding strategy. Our results eliminate this phenomenon and provide more visually satisfying quality of reconstructed frames. Meanwhile, we need less bits for similar visual quality.

 \begin{figure}[h]
 \centering
 \subfigure
 {\includegraphics[scale=0.25]{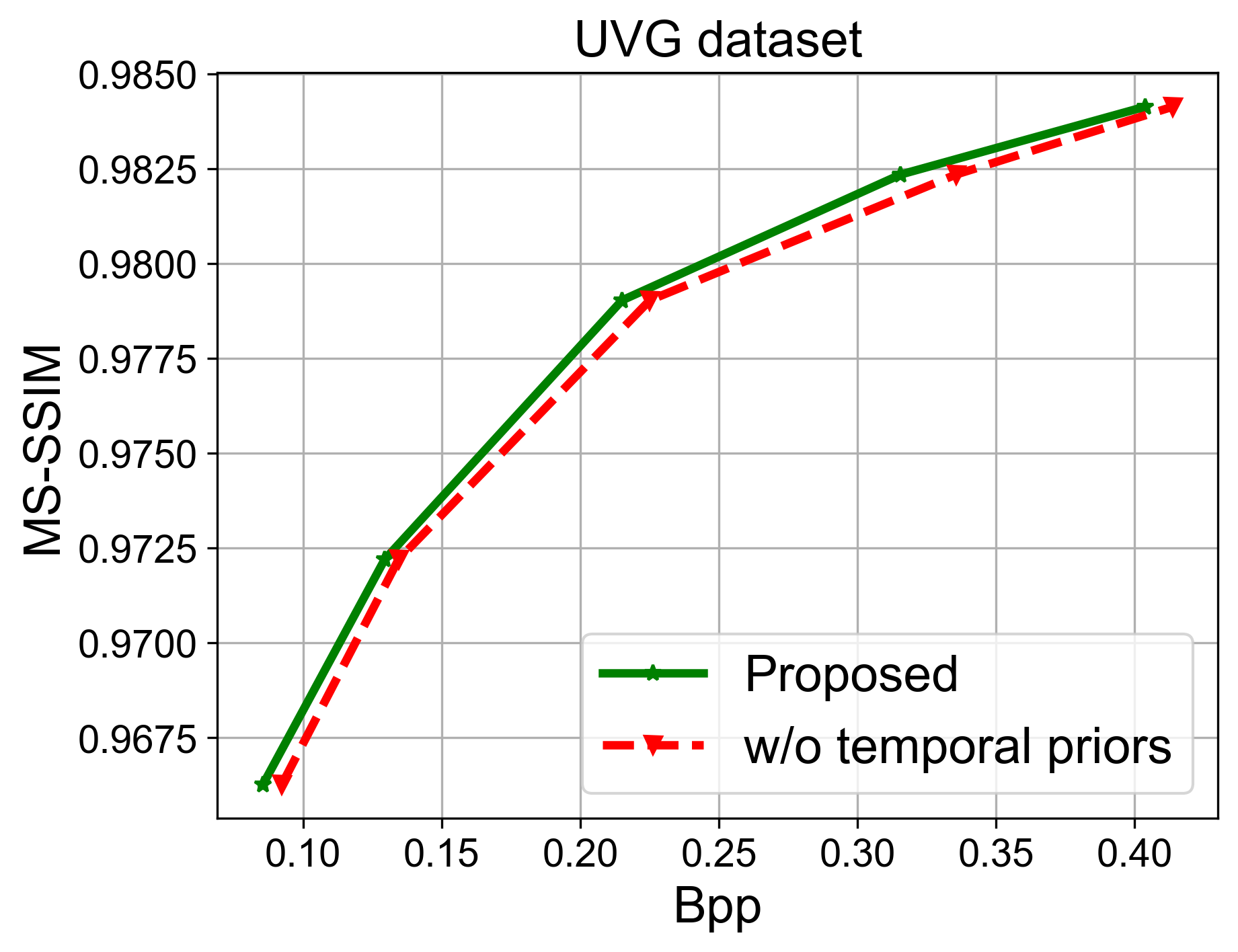}}
 \subfigure{\includegraphics[scale=0.25]{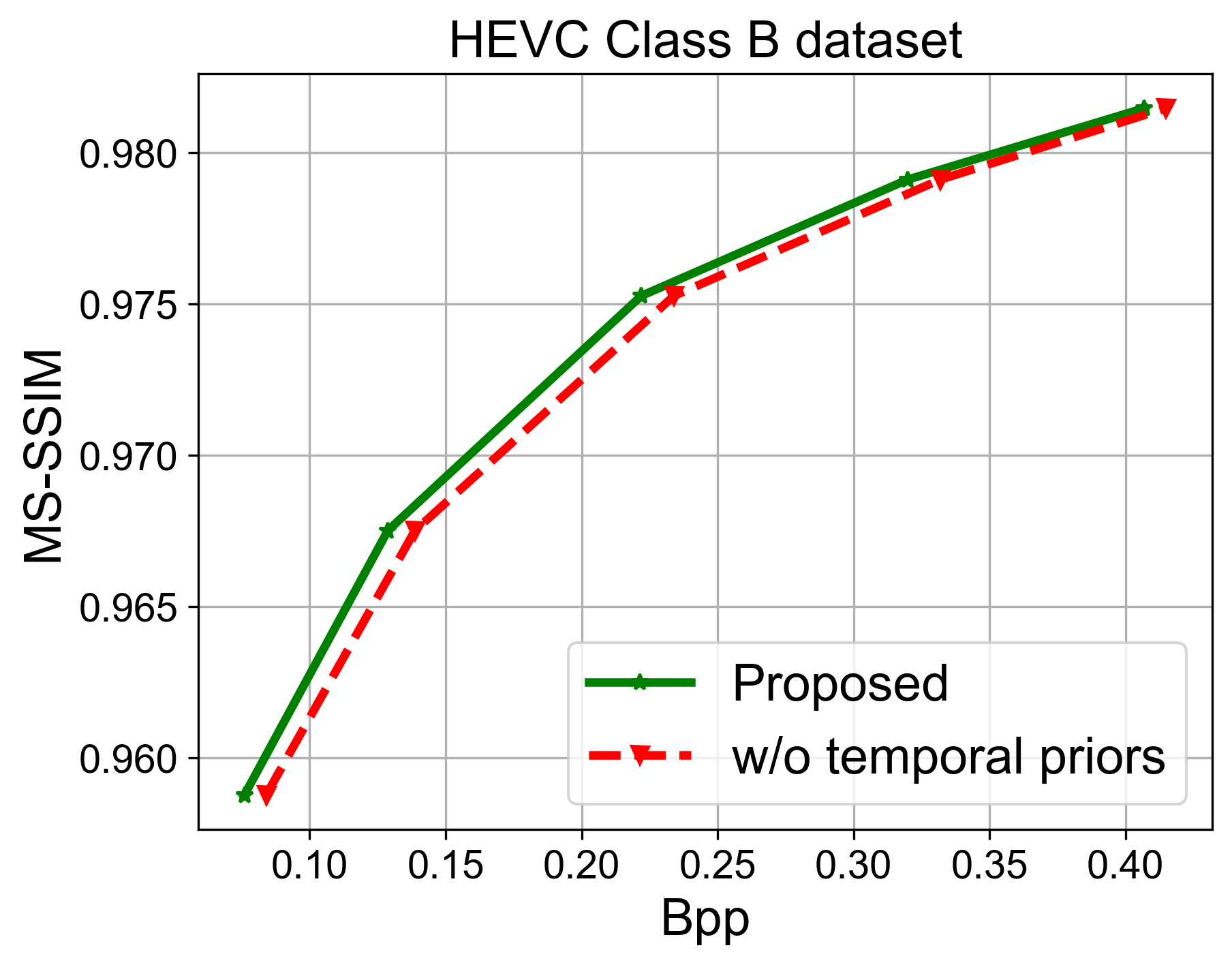}}
 \subfigure{\includegraphics[scale=0.25]{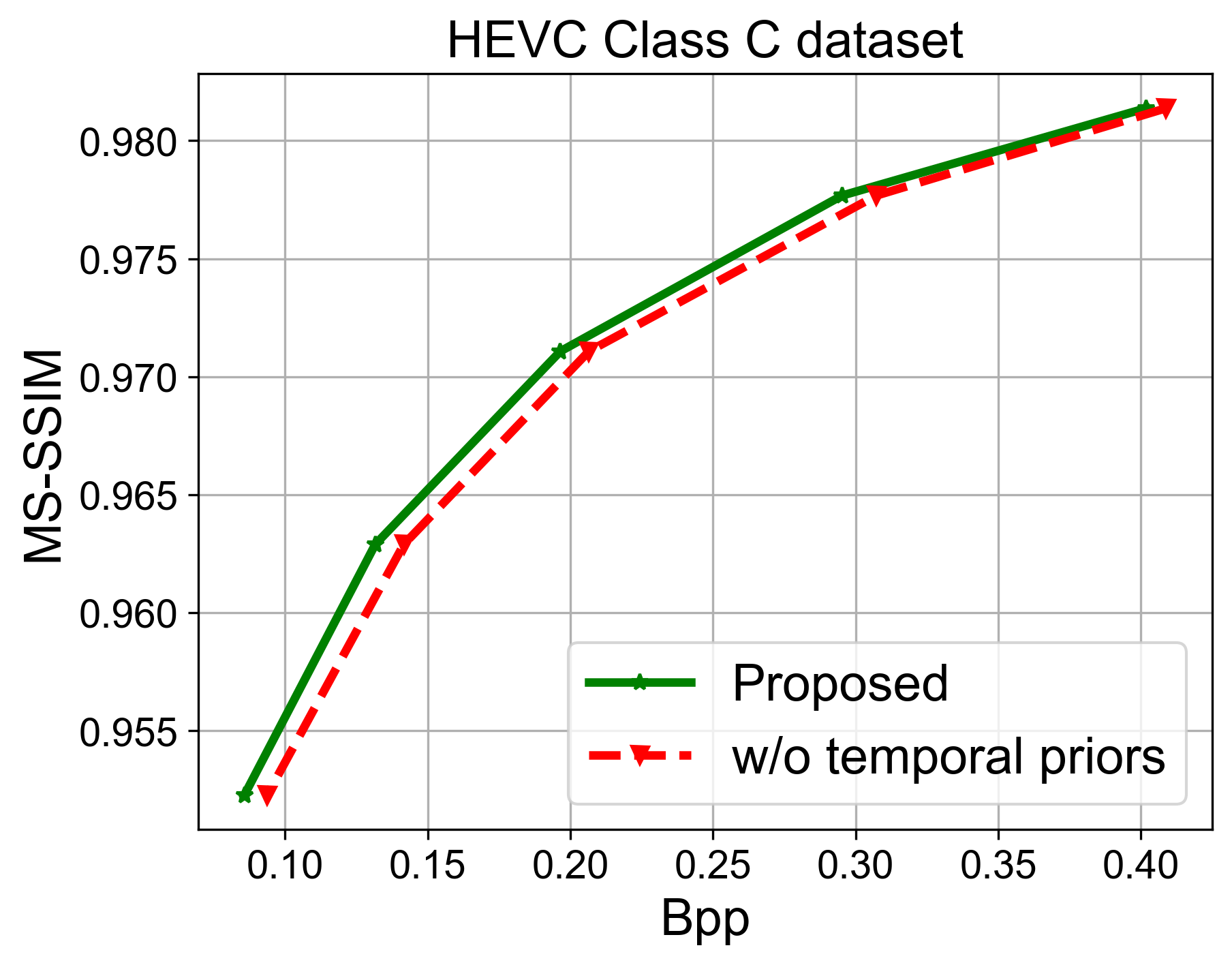}}
 \subfigure{\includegraphics[scale=0.25]{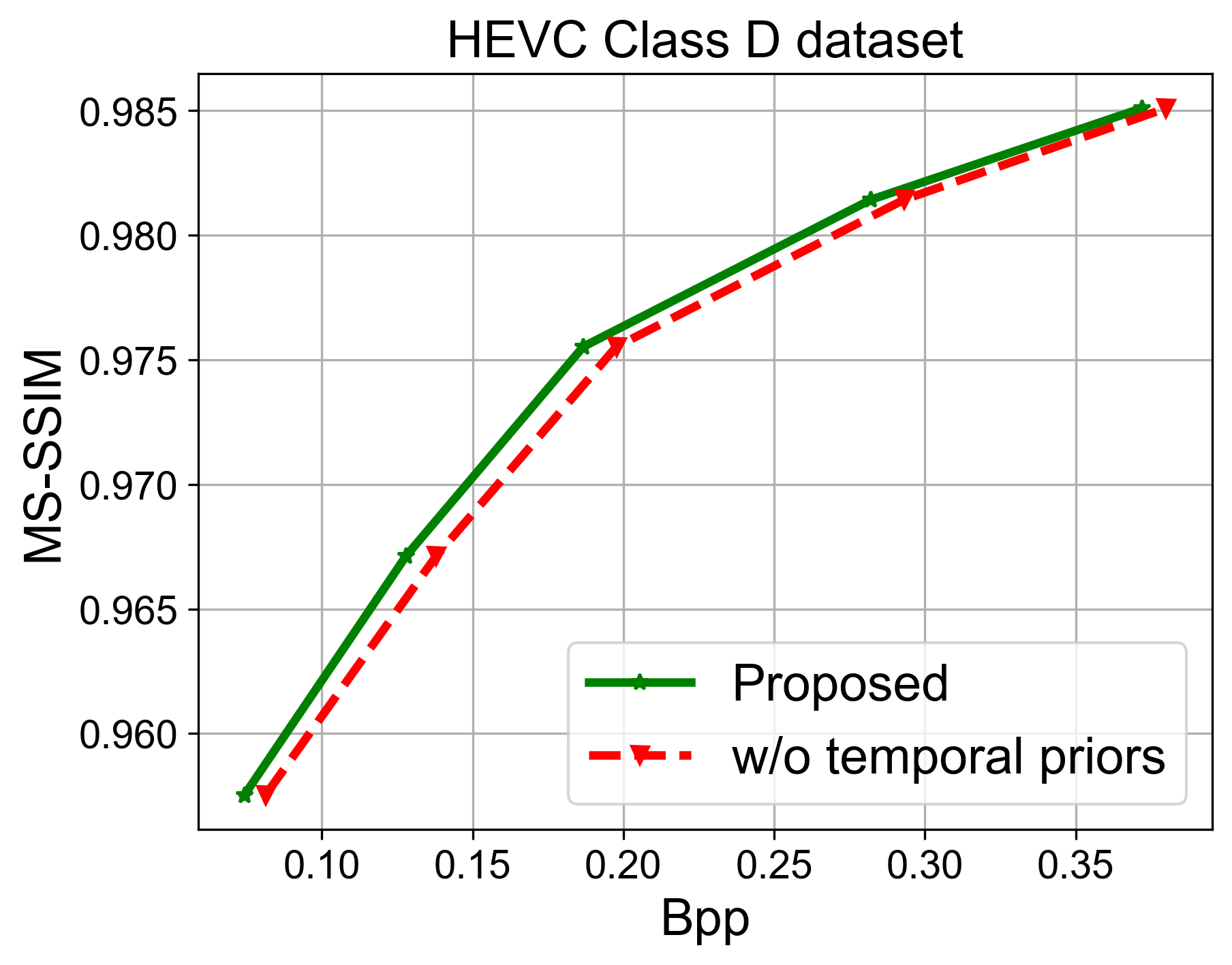}}
 \subfigure{\includegraphics[scale=0.25]{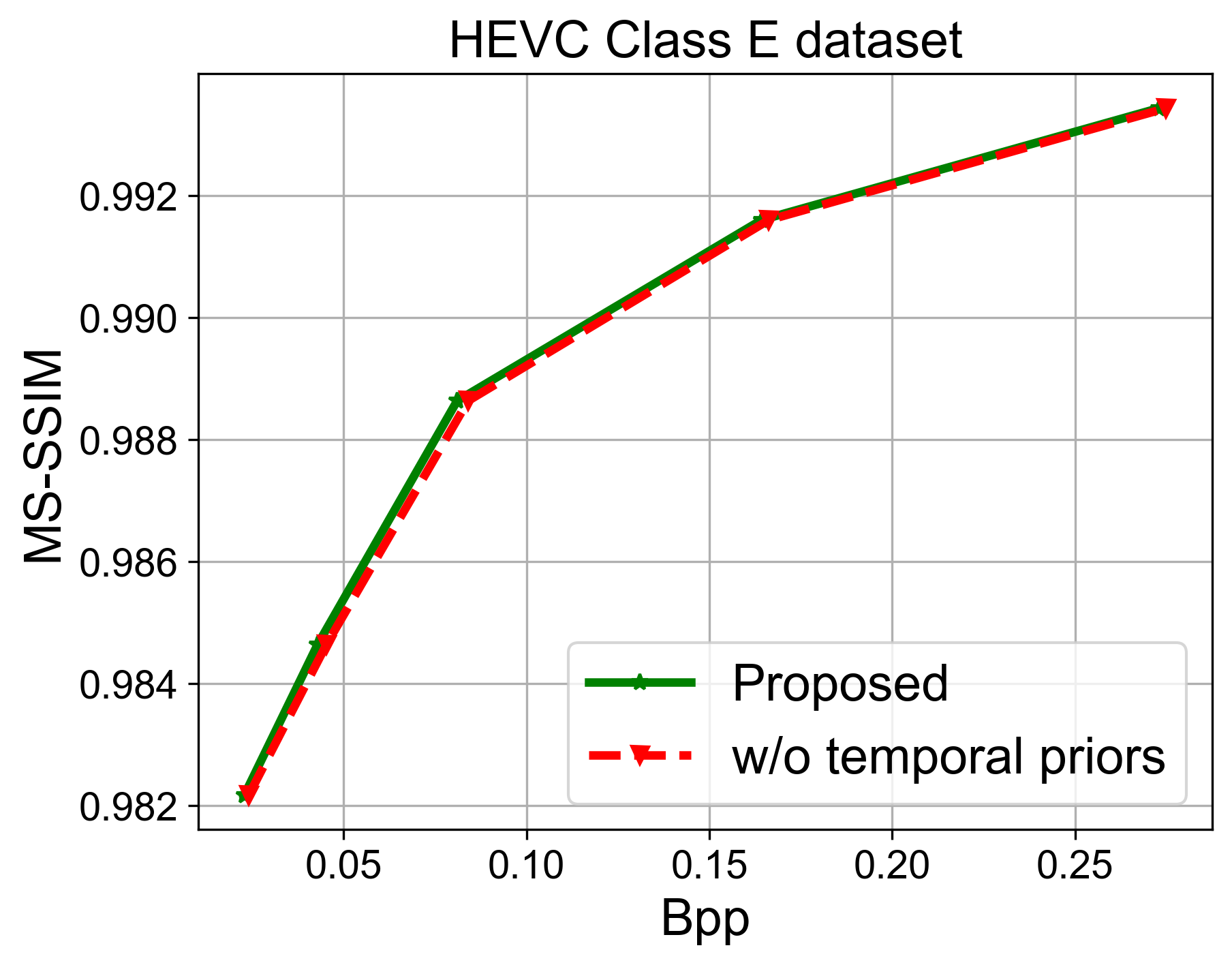}}
 \subfigure{\includegraphics[scale=0.25]{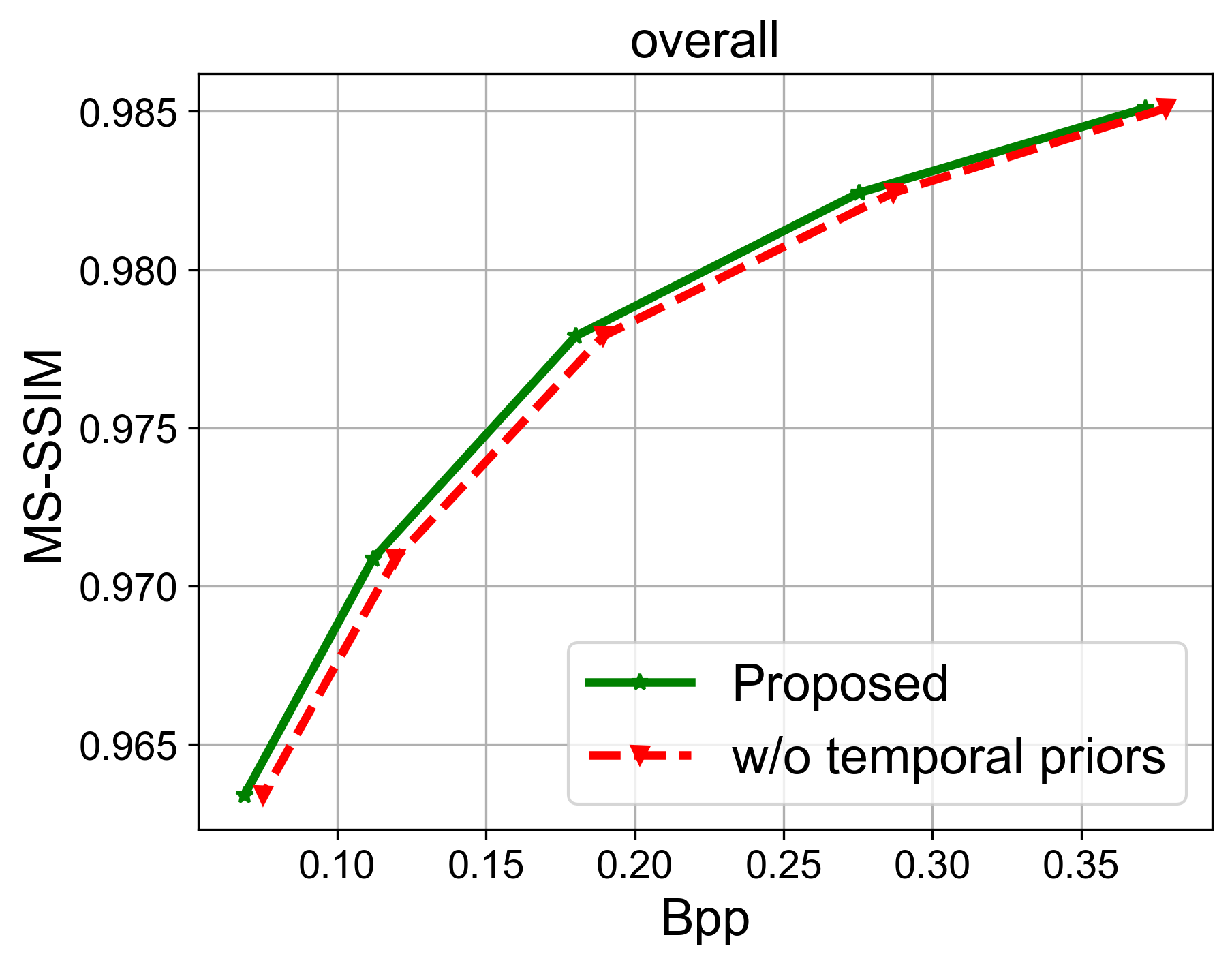}}
 \caption{{\bf Efficiency of Temporal Priors.}  2\% - 10\% loss captured at similar bit rate when removing temporal priors for context modeling.
 }
 \label{rd_abl}
 \end{figure}

\subsection{Ablation Study}
{\bf Non-local Attention Transforms.} Most existing image compressions apply Generalized Divisive Normalization (GDN) as non-linear transform to de-correlate spatial-channel redundancy~\cite{minnen2018joint,balle2018efficient}. Alternative non-local attention transform utilizes NLAM to capture both local and global correlations, leading to the state-of-the-art efficiency as reported in~\cite{liu2019non}. Algorithms in~\cite{minnen2018joint} ranks the second place for coding efficiency. Both methods in~\cite{liu2019non} and~\cite{minnen2018joint} apply the VAE structures.
Fig.~\ref{fig:nlam} experiments the efficiency of NLAM, revealing that performance can be retained closely to ~\cite{minnen2018joint} even by removing all nonlocal operations.



{\bf Second-order Flow Correlations}
In Fig.\ref{fig:redundancies}, we have shown that the second-order motion representations imply the further temporal redundancy between optical flows. Thus,
we present a recurrent state (e.g., ConvLSTM) to aggregate temporal priors for inter coding which can effectively reduce the bits for flow compression.

Temporal priors are fused with autoregressive and hyper priors to improve the context modeling of flow element. Then we use ConvLSTM to combine temporal priors with current quantized features for the updated priors in a recurrent way. Flow prediction provides an effective means to exploit the redundancy between complex motion behaviors.
 Fig.~\ref{rd_abl} shows that efficiency variations when removing the ConvLSTM system from our entire framework without exploiting the temporal correlations, where 2\% to 10\% quality loss is captured.

 Generally, bits consumed by motion information  varies across different content and bit rates, leading to a variety of percentages to the total bits. More bit saving is revealed for low bit rates, and  motion intensive content. For stationary content, such as HEVC Class E, spatial and hyper priors already give a good reference, thus temporal priors are less used.

\section{Conclusions \& Future Work}
In this paper, we present an end-to-end video compression framework and fully exploit the spatial and temporal redundancies. Key novelty laid on the accurate motion representation for exploiting temporal correlation, via both first-order optical flow learning and second-order flow predictive coding. An one-stage unsupervised flow learning is applied with implicit flow representation using quantized features. These features are then compressed using joint spatial-temporal priors by which the probability model is conditioned adaptively.

We evaluate our methods and report the state-of-the-art performances among all the existing video compression approaches, including traditional H.264/AVC, H.265/HEVC, and learning-based DVC. Our approach offers the consistent gains over existing methods across a variety of contents and bit rates.

As for the future study, an interesting topic is to devise implicit flow without actual bits consumption, such as the decoder-side flow derivation, or frame interpolation and extrapolation. Currently, residual shares the same network with intra coding, which may be worth for deep investigation for network simplification. It is also significant to generalize the whole system to more complex video data sets such as spectural video~\cite{cao2016computational} and 3D video~\cite{muller20133d,cao2011semi}.

\section{Acknowledgement}
This work was supported in part by the National Natural Science Foundation of China under Grant 61571215.
\bibliographystyle{aaai}
\bibliography{aaai}

\end{document}